\documentclass[12pt]{iopart}
\renewcommand{\baselinestretch}{0.833}
\def\be{\begin{equation}}
\def\ee{\end{equation}}
\def\bea{\begin{eqnarray}}
\def\eea{\end{eqnarray}}
\def\bean{\begin{eqnarray*}}
\def\eean{\end{eqnarray*}}
\def\thru#1{\mathrel{\mathop{#1\!\!\!\!/}}}

\begin{document}

\title{Exceptional Groups and Physics}
\author{Pierre Ramond}
\address{Institute for Fundamental Theory, Physics Department, University of Florida,\\ Gainesville, FL 32611, USA}
\begin{abstract}
Quarks and leptons  charges  and interactions are derived from gauge theories associated with symmetries. Their space-time labels come from representations of the non-compact algebra of Special Relativity. Common to these descriptions are the Lie groups stemming from their invariances.   Does Nature use Exceptional Groups, the most distinctive among them? We examine the case for and against their use. They do indeed appear in   charge space, as the Standard Model fits naturally inside the exceptional group $E_6$. Further, the advent of the $E_8\times E_8$ Heterotic Superstring theory adds credibility to this venue. On the other hand, their  use as space-time labels  has not been as evident as they link  spinors and tensors under space rotations, which flies in the face of the spin-statistics connection. We discuss a way to circumvent this difficulty in trying to generalize eleven-dimensional supergravity.
\end{abstract}

\section{Introduction}
With the advent of Quantum Mechanics, Lie algebras and the groups they generate have found widespread uses in the description of physical systems.  The  quantum-mechanical state of a particle is determined by labels. Some, like the particle's momentum (or position) are continuous, others like its spin, and charges assume discrete values. All stem from  irreducible unitary representations of Lie algebras. The continuous ones pertain to irreps of non-compact groups, and the discrete ones to compact groups. Mass and spin label the representations of the non-compact group of special relativity the Poincar\'e group, and the color of a quark roam inside a representation of the compact color group $SU(3)$.  Moreover, their interactions are determined by dynamical structures based on these invariance groups. 

Although Nature does not use {\em all}   mathematical structures created by our mathematical friends, it seems to favor some particularly unique and beautiful ones for the description of its inner secrets.  Alas, they often appear in disguised broken-down form, so it is up to us to divine their existence from incomplete evidence: awareness of these structures is an important research tool.

 There are four infinite families of simple Lie algebras,  the garden variety algebras: $A_n\sim SU(n+1)$, $B_n\sim SO(2n+1)$, $C_n\sim Sp(2n)$, and $D_n\sim SO(2n)$, all with $n$ extending to $\infty$.  They describe spacetime rotation, quark and lepton charges, and their associated Yang-Mills gauge structures. Today, $SU(N)$ gauge  theories with $N$ large are intensely studied.  

In the Lie garden, one also finds five rare flowers, the exceptional algebras: $G_2$, $F_4$, $E_6$, $E_7$ and $E_8$, their rank indicated by the subscripts. In view of Nature's fascination with unique structures, they merit further study.    

\section{A Short Course on Exceptional Algebras}
The smallest exceptional algebra~\cite{RAMOND} is $G_2$. It has $14$ parameters and its smallest representation is seven-dimensional, the seven imaginary directions of octonions (Cayley numbers). It is in fact the automorphism group of the octonion algebra. An octonion  $\omega$ is written as

\be
\omega =a^{}_0+a^{}_\alpha\,e^{}_\alpha\  ,~~~~ \alpha=1,2,\dots, 7\ , ~~~~ e^2_\alpha=-1 ,\qquad 
e^{}_\alpha\,e^{}_\beta=\Psi^{}_{\alpha\beta\gamma}\,e^{}_\gamma ,\ee
for $\alpha\ne\beta$, where $\Psi_{\alpha\beta\gamma}$ are totally antisymmetric and equal to $+1$ for the combinations $(\alpha\beta\gamma)=(123)\ ,(246)\ ,(435)\ ,(651)\ ,(572)\ ,(714)\ ,(367)$, and zero otherwise. This algebra is non-associative as their associator does not vanish:

\be[\,e^{}_\alpha,e^{}_\beta,e^{}_\gamma\,]\equiv (e^{}_\alpha\,e^{}_\beta)\,e^{}_\gamma-e^{}_\alpha\,(e^{}_\beta e^{}_\gamma)=2\widetilde\Psi_{\alpha\beta\gamma\delta}\,e^{}_\delta ,\ee
where $\widetilde\Psi_{\alpha\beta\gamma\delta}$ is the dual of the structure constants.  $G_2$ acts on the seven imaginary units. 

There are four Hurwirtz algebras, the real numbers $R$, the complex numbers $C$, the quaternions$Q$,  and the octonions $\Omega$. The three quaternion imaginary units are the Pauli spin matrices (multiplied by $i$), and their automorphism group is $SU(2)$. All have the property that the norm of their product is the product of their norms.  

All other exceptional algebra can be  constructed  terms of $(3\times 3)$ antihermitian traceless matrices with elements over products of two sets of  Hurwitz algebras. This leads to the ``magic square" of Tits and Freudenthal. 
Apply the construction to a  matrix with elements over $\Omega\times\Omega'$. An  ``octonionic octonion" has $8\times 8=64$ elements, while an imaginary one has $7+7=14$ elements. In an antihermitian traceless matrix, this accounts for $3\times 64+2\times 14=220$ parameters. Adding to them the two automorphism groups, we get the $248$ parameters of $E_8$, the largest exceptional Lie algebra . If Lie algebras can be associated with cars, surely $E_8$ is the Delahaye of Lie algebras!

\section{Charge Spaces}
The state of an elementary particle is labelled at a given time as

\be
\vert\, m\, ,s\, , x^i\, ,x^-\,, \{s_a\}\ ;\, \xi_1 ,\xi_2 ,\dots\, \xi_N \,>_t\ ,\ee 
where the first set are the space-time labels given in light-cone coordinates: the continuous transverse positions $x^i$, where $i$ runs over the transverse dimensions of space, the spins $s$ (more than one in higher dimensions). The second labels are the internal charges $\xi_\alpha$ which are described by irreps of compact Lie algebras. The space-time is thus written in terms of orthogonal group of rotations in the transverse space, subgroup of  the semi-simple non-compact Poincar\'e group. On the other hand, the discrete internal charges belong to representations of compact simple Lie groups. 

Quarks and lepton charges span representations of the Standard Model group $SU(3)\times SU(2)\times U(1)$. Remarkably they fit snuggly into two representations of the larger $SU(5)$~\cite{GG}.

\be SU(5)\supset  SU(3)\times SU(2)\times U(1)\ ,\ee
with three families transforming as ${\bf \bar 5}\oplus {\bf 10}$. 
 With the discovery of neutrino masses it is almost certain  that each neutrino has a 
Dirac partner, the right-handed neutrino.  With it, each family fits in the  fundamental spinor  representation of $SO(10)$~\cite{FM}:

\be
SO(10)\supset SU(5)\times U(1)\ ;\qquad {\bf 16}={\bf \bar 5}\oplus {\bf 10}\oplus {\bf 1}\ .\ee
It is amazing that the natural algebra with one rank higher is that of the exceptional $E_6$~\cite{ESIX}, with 

\be
E_6\supset SO(10)\times U(1)\ ;\qquad {\bf 27}={\bf 16}\oplus {\bf 10}\oplus {\bf 1}\ ,\ee
which is a complex representation. $E_6$ and the spin representations of orthogonal groups $SO(4n+2)$, $n\ge 2$ are the only fundamental  complex representations with no anomalies.  This of course open the road to $E_8$:

\be
E_8\supset E_7\times SU(2)\supset E_6\times U(1)\ .\ee
This ladder to exceptional algebras is even more apparent through their Dynkin diagrams.

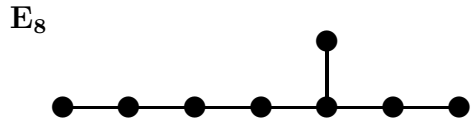
\begin{picture}(200,100)(-100,20)
\put(5,80){\shortstack{$\bf E_8$}}
\thicklines
\put(25,50){\circle*{8}}
\put(50,50){\circle*{8}}
\put(75,50){\circle*{8}}
\put(100,50){\circle*{8}}
\put(125,50){\circle*{8}}
\put(150,50){\circle*{8}}
\put(175,50){\circle*{8}}
\put(125,75){\circle*{8}}
\put(25,50){\line(150,0){150}}
\put(125,50){\line(0,25){25}}
\end{picture}

\noindent By chopping-off one dot at a time, one arrives at the Dynkin of the Standard Model

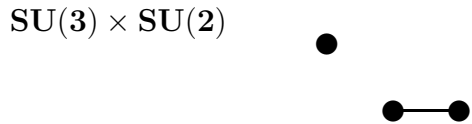
\begin{picture}(200,100)(-100,20)
\put(5,80){\shortstack{$\bf SU(3)\times SU(2)$}}
\thicklines
\put(150,50){\circle*{8}}
\put(175,50){\circle*{8}}
\put(125,75){\circle*{8}}
\put(150,50){\line(25,0){25}}
\end{picture}

Exceptional groups make their appearance in superstring theory. The gauge group of the most promising heterotic string~\cite{HET}  in ten space-time dimensions is nothing but $E_8\times E_8$ with $496$ gauge parameters (496 is a perfect number: can anyone doubt string theory~\cite{BOYA}?). There, one  compactifies over a six-dimensional manifold to get to four space-time dimensions. To preserve supersymmetry, the manifold must have $SU(3)$ holonomy. A trip in the extra dimensions gets you back where you started modulo $SU(3)$ , and this is compensated  by the $SU(3)$ obtained from $E_8\supset E_6\times SU(3)$. Thus $E_6$ is naturally obtained!  The number of families is the number of holes in the six-dimensional Calabi-Yau manifold~\cite{CANDELAS}. 

\section{Space Charges}
 Exceptional groups naturally contain orthogonal groups as subgroups

\bea E_8&\supset &SO(16)\ ;\qquad~~~~~~~~~~ E_7\supset SO(12)\times SO(3)\  ;\cr
E_6&\supset & SO(10)\times SO(2)\ ;~~~~ ~F_4~\supset~ SO(9)\ , ~~~~ G_2~\supset~ SO(3)\times SO(3)\ .\eea 
They could therefore contain (in their non-compact form) the conformal group in $D$ spacetime dimensions$SO(D,2)$ and its Poincar\'e subgroup or else as contracted form of the above.   However any role that exceptional groups may play in the description of space charges  
 (position, mass,spin,\dots) has to be quite subtle. 

The reason is that their representations contain both spinorial and tensorial representations of their orthogonal subgroups. For instance,   the fundamental irrep of $F_4$ 
$${\bf 26}~=~{\bf 16}\oplus {\bf 9}\oplus {\bf 1}\ ,$$  
  contains both the $SO(9)$ spinor and the vector representations. Thus  $F_4$ transformations naturally mix these, but in quantum theory this is like mixing apples and oranges: space spinors obey Fermi-Dirac statistics while the vectors are Bose-Einstein. This simple fact makes their relevence to space charges indirect to say the least. On the other hand, fermions and bosons do coexist in Nature and there must be some symmetry which links them. It is well at this point to examine the difference between bosons and fermions

\subsection{Fermion-Boson Confusion}
In four dimensions, fermions and bosons are naturally differentiated,  as  fermions have half-odd integer helicities while the boson helicities are integers.  In $d+1$ spacetime dimensions, fermions transform as spin representations of the transverse little group $SO(d-1)$, while bosons are transverse tensors.  As a result in most dimensions, fermions and bosons have different dimensionalities, but there are exceptions: in $1+1$ dimensions, there is no transverse little group and both fermions and bosons are uni-dimensional. This makes it easier to confuse them and there is the well-know phenomenon of bosonisation or fermionisation. 

In $9+1$ dimensions, the little group is $SO(8)$, with its unique triality property according to which bosons and fermions are group-theoretically equivalent, and this is the domain of superstring theories where this triality is put to excellent use. The Dynkin diagram of $SO(8)$ displays this triality 

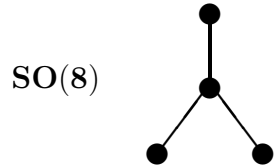
\begin{picture}(200,100)(-100,20)
\linethickness{1.2pt}
\put(50,75){\shortstack{$\bf SO(8)$}}
\put(125,103){\circle*{8}}
\put(105,50){\circle*{8}}
\put(145,50){\circle*{8}}
\put(125,75){\circle*{8}}
\thicklines
\put(125,75){\line(0,25){25}}
\put(105,50){\line(3,4){18}}
\put(145,50){\line(-3,4){18}}
\end{picture}

\noindent and it is the Mercedes of Lie groups. This triality is explicit in the $F_4\supset SO(8)$ decomposition. One of the great surprises in string theories has been the emergence of a new theory which contains all string theories; it is called M-theory and it is not a string theory and lives in one more space dimension than the superstrings. The heterotic string theory can be obtained by compactifying M-theory over the line $S^1/Z_2$. The infrared limit of M-theory is $N=1$ supergravity in eleven dimensions~\cite{ELEVEN}.  When compactified on a $d$-torus, one finds a non-compact exceptional group, $E_{d(d)}$, where the number in parenthesis is the number of non-compact generators minus the number of compact ones. In particular for $d=8$ one obtains a theory in $2+1$ dimensions with a non-compact version of $E_8$~\cite{NICOLAI}.

It seems that there is also a special arrangement between fermions and bosons in eleven spacetime dimensions. Yet there is nothing remarkable about the transverse little group $SO(9)$. Its Dynkin diagram

\begin{picture}(200,100)(-100,20)
\linethickness{1.2pt}
\put(100,75){\shortstack{$\bf SO(9)$}}
\put(80,50){\circle*{8}}
\put(105,50){\circle*{8}}
\put(130,50){\circle*{8}}
\put(155,50){\circle{8}}
\put(84,50){\line(18,0){18}}
\put(109,50){\line(18,0){18}}
\put(132,53){\line(22,0){21}}
\put(132,47){\line(22,0){21}}
\end{picture}

\noindent does not display any symmetries; it is more like a Trabant than a Mercedes or a Delahaye! How can there be any confusion between bosons and fermions?  Yet it describes the space in which M-theory roams! Finally we note two interesting anomalous Dynkin embeddings which might have hitherto unsuspected applications. The first is 

$$SO(16)~\supset~ SO(9)\ ,$$
in which the sixteen-dimensional {\em spinor} representation of $SO(9)$ fits snuggly into the sixteen-dimensional {\em vector} representation of $SO(16)$. We will come back to it later. The second is 

$$SO(26)~\supset~F_4\ ,$$
which equates the $26$-dimensional vector irrep of $SO(26)$ to that of $F_4$. Its real form

$$SO(25,1)~\supset~F_{4(-20)}~\supset~ SO(9)\ ,$$
 may provide a heterotic path to eleven dimensions starting from the original bosonic theory.

\section{$N=1$ Supergravity in $11$ Dimensions}
Supergravity in eleven spacetime dimensions is the infrared limit of M-theory. It is a local field theory,  believed to diverge at three loops. On the light-cone, the theory is described by a chiral superfield with $256$ components

$$\Phi(y^-,\vec x\ ,\theta^\alpha)~=~\phi(y^-,\vec x)+\theta^\alpha\,\psi_\alpha(y^-,\vec x)+\cdots \theta^1\theta^2\cdots\theta^8\lambda(y^-,\vec x)\ ,$$
expanded in terms of eight complex anticommuting Grassman variables, and where  $y^-$ is the displaced chiral coordinate

$$y^-~=~x^- -i\bar \theta\,\theta/\sqrt{2}\ .$$
Introduce the sixteen $(256\times 256)$ Dirac matrices

$$\{\,\Gamma^a,\Gamma^b\,\}~=~2\delta^{ab}\ ,\qquad a,b=1,2\dots 16\ ,$$
with vector indices transforming as the $SO(9)$  spinor(recall the anomalous Dynkin embedding).  These are not to be confused with the $(16\times 16)$ nine Dirac matrices which transform as $SO(9)$ vectors

$$\{\,\gamma^i,\gamma^j\,\}~=~2\delta^{ij}\ ,\qquad i,j=1,2\dots ,9\ .$$
Together they allow for a neat way of writing the $SO(9)$ generators acting on this superfield

$$S^{ij}~=~-\frac{i}{4}\,\Big(\gamma^{ij}\Big)_{ab}\Gamma^{ab}\ ,$$
where in the usual notation $\gamma^{ij}=\gamma^i\gamma^j$, $i\ne j$, $\Gamma^{ab}=\Gamma^a\Gamma^b$, $a\ne b$. The $52$ $F_4$ parameters i split into  the $36$  $S^{ij}$ which generate $SO(9)$, and  sixteen $SO(9)$ spinors, $T^a$. Algebraic closure is given by

$$[\,T^a,T^b\,]~=~\frac{i}{2}\,\Big(\gamma^{ij}\Big)^{a}\,S^{ij}\ ,$$
so there is a  whiff of $F_4$ in the light-cone description of $N=1$ SUGRA in eleven dimensions. We shall see later this is the tip of a beautiful algebraic structure.  Their action on the superfield show it to split in three $SO(8)$ representations, the $\bf 44$ of the symmetric second rank traceless tensor, the $\bf 84$ the antisymmetric third rank tensor, and $\bf 128$ the Rarita-Schwinger spinor-vector field.    It is convenient to write the superfield in terms of the three highest weight components of each representation (in our basis)

$$\Phi(y^-,\vec x\ ,\theta^\alpha)~=~\theta^1\theta^8\Big( h(y^-,\vec x)+\theta^4\,\psi(y^-,\vec x)+ \theta^4\theta^5\,A(y^-,\vec x)\Big)\ ,$$
where $h\ ,\psi\ ,A$ are  the highest weights of the $\bf 44$, $\bf 128$, and $\bf 84$, respectively.  Define the Dynkin indices as

$$I^{(k)}~=~\sum_{\rm irrep}\,w^k\ ,$$
in terms of the length of each weight $w$. They satisfy the composition law

$$I^{(k)}_{\bf r\times s}~=~I^{(0)}_{\bf r}\,I^{(k)}_{\bf  s}+I^{(0)}_{\bf  s}\,I^{(k)}_{\bf r}\ ,$$
where $ \bf r$ and $\bf s $ are any two irreps. $I^{(0)}$ is the dimension of the representation. Since $SO(9)$ has rank four it has four such independent indices. The three SUGRA irreps have much in common, as the following table shows
\hskip 2cm
\begin{center}
\begin{tabular}{|c|c|c|c|}
\hline
$~{\rm irrep}~$& $(1001)$&$ (2000)$ & $(0010)$   \\
 \hline \hline         
$~D~ $&$  128$ & $ 44$ & $ 84$  \\
 \hline    
$~I_2~$& $256$& $88$ & $168$  \\
\hline
$~I_4~$& $640$& $232$ & $408$ \\                                                            
 \hline
$~I_6~$&$1792$& $712$ &$1080$\\ 
\hline
$~I_8~$&$5248$& $2440$ &$3000$\\ 
\hline
 \end{tabular}\end{center}
\vskip 0.3cm
It shows that

$$I^{(k)}_{\bf 128}~=~I^{(k)}_{\bf 44}+I^{(k)}_{\bf 84}\ ,\qquad k=0,2,4,6\ ,$$
but the sum rule fails for the higher invariant $k=8$. It has been conjectured~\cite{CURTRIGHT} that it is this failure that is responsible for the non-renormalizability of $N=1$ SUGRA. 

Amazingly,  this pattern of equalities is repeated for an infinite number of sets of three $SO(9)$ representations, which  describe higher spin massless particles~\cite{PR}. It has to do with the fact that there are three equivalent ways to embed $SO(9)$ inside $F_4$~\cite{GKRS}. This is the octonionic equivalent I-spin, U-spin and V-spin which label three equivalent ways to embed $SU(2)$ inside $SU(3)$. The $F_4$ Weyl chamber is 1/3 that of $SO(9)$. Take a highest weight in the $F_4$ Weyl chamber, $\lambda$. Let $\rho$ be the sum of the fundamental weights. There exist two Weyl reflections $C$, which map $\lambda$ outside the $F_4$ Weyl chamber, but stay inside that of $SO(9)$.   Hence there is a unique way to associate one $F_4$ representation to three $SO(9)$ irreps. The mapping is 

$$C\bullet\lambda~=~C\,(\lambda+\rho_{F_4})-\rho_{SO(9)}\ .$$
This mapping associates with each $F_4$ irrep, a set of  three $SO(9)$ representations
 called Euler triplets.  Equality betwen its Dynkin indices is guaranteed by the character formula

$$V_\lambda\otimes S^+-V_\lambda\otimes S^-~=~\sum_C\,sgn(C)\,{\cal U}_{C\bullet\lambda}\ ,$$
where $V_\lambda$ is any $F_4$  representation written in terms of its $SO(9)$ content, $S^\pm$ are the two spinor irreps of $SO(16)$ also written in terms of $SO(9)$ through the anomalous Dynkin embedding

$${\bf 128}~=~{\bf 44}\oplus{\bf 84}\ ;\qquad {\bf 128}~=~{\bf 128}\ .$$
The failure of the equality for the eigth order invariant is linked to the fact that $S^+$ and $S^-$ have different Pfaffian invariants~\cite{ED}.  
One recognizes the ``trivial" Euler triplet as the three fields of $N=1$ SUGRA in eleven dimensions associated with $\lambda=0$. 
This character formula is akin to an index formula for Kostant's operator associated with the coset $F_4/SO(9)$, the sixteen-dimensional projective Cayley-Moufang plane. Euler triplets are solutions of Kostant's equation~\cite{KOS}

$$\thru{\cal K}\,\Psi~\equiv~\Gamma^a\,T^a\,\Psi~=~0\ ,$$
where the $T^a$ generate the $F_4/SO(9)$ tranformations. 

$$
[\,T_{}^a\,,\,T_{}^b\,]~=~i\,f^{[ij]\,ab}_{}\,T^{ij}_{}\ .$$
Kostant's operator commutes with the generalized $SO(9)$ generator made up of an ``orbital" and the previously defined  ``spin" part

$$
L_{}^{ij}~\equiv~T_{}^{ij}+S_{}^{ij}\ .$$
The solutions to Kostant's equation are the Euler triplets, and the trivial solution is the SUGRA triplet. The number of representations in each Euler set is the ratio of the order of the  $F_4$ and $SO(9)$ Weyl groups. It is also the Euler number of the coset manifold, hence the name. 

It is convenient~\cite{FULTON} to express the $F_4$ in terms of three sets of $26$ real coordinates: $u^{}_i$ which transform as transverse space vectors, $u^{}_0$ as scalars, and $\zeta^{}_a $ as space spinors. This enables us to write the Euler triplets as chiral superfields of the form~\cite{BRX}

$$\Phi(y^-,\vec x\ ,\theta^\alpha)~=~\theta^1\theta^8\Big( h(y^-,\vec x\ ,u_i\ ,\zeta_a)+\theta^4\,\psi(y^-,\vec x)+ \theta^4\theta^5\,A(y^-,\vec x)\Big)\ ,$$
where now the components $h$, $\psi$ and $A$ are the highest weight components of the three irreps with definite polynomial dependence on the new coordinates. For the proper spin-statistics interpretation, the twistor-like variables $\zeta_a$ must appear quadratically. It turns out that the $\zeta$'s appear in even powers only for those Euler triplets that have the same number of bosons and fermions!

The physical interpretation of these triplets is still unclear. Their  quantum numbers suggest that they can be related by the emission of fields with specified quantum number, in analogy with the transition within a gauge multiplet by emission of a $W$-boson. In particular one recognizes the two-form field, so it is possible that emission of a two-form potential from the superparticle in eleven dimension might generate the other triplets, but supersymmetry is broken in the process.

Poincar\'e invariance requires the Euler triplets to be massless as there are not enough fields among them to complete into massive $SO(10)$ little group multiplets. Furthermore there is no supersymmetry relating members of an Euler triplet except, of course, the first one. There are grave difficulties~\cite{DIFFICULT} when coupling to gravity a massless particle with spin greater than two in flat space-time~\cite{VASILIEV}: either one gets a relativistic theory ghosts or else a theory that does not satisfy Lorentz invariance. There are no such objections with an infinite number of such particles, which would correspond to a highly non-local theory.

There are indications that one may need in fact an infinite number of Euler triplets. If the divergences of supergravity are linked to the lack of cancellation in $I^{(8)}$, the same would be true for any Euler triplet contribution to a loop amplitude, but the sign of the deficit is the same for all triplets as it is proportional to the dimension of the $F_4$ representation from which it originates. To get a cancellation with manifestly positive quantities, an infinite number are required, in the sense of $\zeta$-function regulari$\zeta$ation. The dimension of any $F_4$ representation is a $24$th order polynomial in the Dynkin integer indices, and the $\zeta$-function of even order vanish, so there might be hope. However it is clear that this is not the language to address this issue as there are formidable technical difficulties to overcome before being able to carry out this program. 

\section{Exceptional Jordan Algebra}
This tour ofExceptional groups in Physics would not be complete without a mention of the Exceptional Jordan Algebra~\cite{FEZA}. Jordan algebras  provide   an alternate way of describing Quantum
Mechanics in terms of its observables. Let    $J_a$ be any observable, we introduce the commutative product
\be
J_a\circ J_b~=~J_b\circ J_a\ ,
\ee
 which maps observables into observables. Since matrices do not commute, the Jordan associator 
\be(J_a,J_b,J_c)~\equiv ~J_a\circ (J_b\circ J_c)- (J_a\circ J_b)\circ J_c\ .
\ee
 is not  zero, but satisfies the Jordan identity
 \be(J^{}_a,J^{}_b,J^2_a)= 0 \ .
\ee
These equations   serve as the postulates of the 
commutative but non-associative Jordan Algebras.By writing  the Hamiltonian in terms of two hermitian matrices
\be
H={i\over 4}[A,B]\ ,\ee
we can express time evolution in terms of the Jordan associator
\be
i\hbar\frac{\partial J}{\partial t}=(A,J,B)\ .\ee
There would be nothing new in this rewriting of  Quantum Mechanics, if it were not for  Jordan, von Neumann and Wigner~\cite{JVW} who noticed that the Jordan axioms were satisfied by  $(3\times 3)$ hermitian
matrices over octonions. The non-associativity  of the octonion forbids a Hilbert space interpretation, and this is what makes it special. It is   known as the exceptional Jordan algebra (EJA). Its group of derivations of the EJA is nothing but the exceptional $F_4$! 
G\"ursey suggested  the EJA  as  label for internal charges, especially since  $F_4\supset
SU(3)\times SU(3)$. Our analysis with Euler triplets suggest rather that   the $SO(9)$ subgroup be  interpreted as the light-cone little group in eleven dimensions~\cite{FRANCQUI}. If  $SO(9)$ is the light-cone little group in eleven dimensions, we want time evolution to preserve it. To that effect,   we need to couple the EJA to an external  field that transforms  non-trivially under $SO(9)$. Otherwise, time evolution with a fixed external potential preserves at most $SO(7)$.  If the  $SO(9)$ subgroup of EJA automorphism group $F_4$ can indeed be identified with the light-cone little group in eleven space-time dimensions, it will suggest the EJA as the charge space of a very special system.

\end{document}